\documentclass[12pt,prd,aps,epsfig,floats,preprint,eqsecnum,nofootinbib]{revtex4}
\usepackage[hypertex]{hyperref}
\usepackage{graphicx,amsmath,amssymb,epsfig,verbatim,mathrsfs,array,layout,textcomp,latexsym}
\usepackage{multirow}

\newcommand{\be}{\begin{equation}}
\newcommand{\ee}{\end{equation}}
\newcommand{\bea}{\begin{eqnarray}}
\newcommand{\eea}{\end{eqnarray}}
\newcommand{\beq}{\begin{equation}}
\newcommand{\eeq}{\end{equation}}
\newcommand{\beqa}{\begin{eqnarray}}
\newcommand{\eeqa}{\end{eqnarray}}
\newcommand{\no}{\nonumber}
\def\lsim{\mathrel{\rlap{\lower4pt\hbox{\hskip1pt$\sim$}}
     \raise1pt\hbox{$<$}}}         
\def\gsim{\mathrel{\rlap{\lower4pt\hbox{\hskip1pt$\sim$}}
     \raise1pt\hbox{$>$}}}         
\begin{document}

\hbox{DO-TH-10/01}

\title{Flavor in Supersymmetry: Anarchy versus Structure}

\author{Gudrun Hiller}\email{ghiller@physik.uni-dortmund.de}
\affiliation{Institut f\"ur Physik, Technische Universit\"at Dortmund,
   D-44221 Dortmund, Germany}

\author{Yonit Hochberg and Yosef Nir\footnote{The Amos de-Shalit chair of theoretical
     physics}}\email{yonit.hochberg,yosef.nir@weizmann.ac.il}
\affiliation{Department of Particle Physics and Astrophysics,
   Weizmann Institute of Science, Rehovot 76100, Israel\vspace*{1cm}}

\vspace*{1cm}
\begin{abstract}
Future high-precision flavor experiments may discover a pattern of
deviations from the standard model predictions for flavor-changing
neutral current processes. One of the interesting questions that can
be answered then will be whether the flavor structure of the new
physics is related to that of the standard model or not. We analyze
this aspect of flavor physics within a specific framework:
supersymmetric models where the soft breaking terms are dominated by
gauge-mediation but get non-negligible contributions from
gravity-mediation. We compare the possible patterns of non-minimally
flavor-violating effects that arise if the gravity-mediated
contributions are anarchical {\it vs.} the case that they are
structured by a Froggatt-Nielsen symmetry. We show that combining
information on flavor and CP violation from meson mixing and electric
dipole moments is indicative for the flavor structure of
gravity-mediation.
\end{abstract}
\maketitle

\section{Introduction}\label{sec:intro}
The program of high $p_T$ experiments at the Tevatron and at the LHC,
and the program of low energy flavor machines, such as present and
future B-factories and the LHCb, are complimentary to each other. On
one hand, understanding the flavor structure of TeV scale new physics
is likely to shed light on the underlying theory at much higher energy
scales, perhaps as high as the Planck scale. On the other hand,
measuring new flavor parameters may lead to progress in understanding
the flavor structure of the standard model itself, namely the
smallness and hierarchy that appear in the Yukawa couplings. In this
work, we explore this complementarity in a specific new physics
framework, and provide a concrete demonstration of how high precision
flavor measurements will lead to progress in understanding both the
new physics and the standard model flavor puzzles.

The consistency of all measurements of flavor-changing neutral current
(FCNC) processes with the standard model predictions requires that the
flavor structure of new physics at the TeV scale is highly nontrivial.
In particular, it seems likely that this flavor structure should be
closely related to the flavor structure of the standard model Yukawa
couplings. The most extreme application of this hint from low energy
flavor measurements is the assumption of minimal flavor violation (MFV)
\cite{D'Ambrosio:2002ex,Hall:1990ac,Chivukula:1987py,Buras:2000dm,Kagan:2009bn}.
The MFV principle states that the only source of flavor violation,
even in interactions involving new particles, are the Yukawa matrices
of the standard model. However, while the flavor constraints suggest
that the dominant flavor structure of new physics should be MFV, there
is certainly room for sub-dominant contributions that are not MFV. The
discovery of such non-MFV physics will be of utmost interest. Thus,
the first questions that future flavor measurements should explore,
relevant to the new physics flavor puzzle, are the following:

\begin{itemize}
\item Are there non-MFV effects in the new physics? At what level do
  they appear?
\end{itemize}

If, indeed, non-MFV interactions are established, then finding out
their pattern would be of much interest. Indeed, there is a wealth of
possible FCNC sectors. In the quark sector alone, there are six
different relevant transitions: $s\to d$, $c\to u$, $b\to d$, $b\to
s$, $t\to c$ and $t\to u$. Three of these -- the $b$ and $c$ decays --
can be better measured in the B-factories and in LHCb, while the $t$
decays can be explored at the LHC. Understanding this pattern will
allow us to answer yet another question, relevant to the standard
model flavor puzzle:

\begin{itemize}
\item Is the flavor pattern of the non-MFV new physics related to the
  standard model flavor pattern or not?
\end{itemize}

Having these questions in mind, we focus in this work on
supersymmetric (SUSY) models with a hybrid gauge- and
gravity-mediation of supersymmetry breaking \cite{ArkaniHamed:1997km}.
Gauge-mediation is well motivated since it solves the flavor problem
of generic supersymmetric models. It should be kept in mind, however,
that, in principle, gravity-mediated contributions are unavoidable.
What is called {\it pure gauge-mediation} has an implicit assumption
that the gravity-mediated contributions are quantitatively negligible.
Indeed, this is the case if the source of supersymmetry breaking are
$F$-terms with scales that are many orders of magnitude below $(m_Z
m_{\rm Pl})$. Pure gauge-mediation, if realized in Nature, will
not provide us with additional data to try and understand the standard
model flavor puzzle, namely the physics that leads to the structure
observed in the Yukawa sector.

In contrast, in the framework called {\it hybrid gauge- and
  gravity-mediation}, there {\it is} an $F$-term within a few orders
of magnitude below $(m_Z m_{\rm Pl})$. This class of models provides an
example of a well-motivated theoretical framework where the dominant
flavor structure of the new physics (the soft supersymmetry breaking
parameters) is MFV, coming from gauge-mediation, but there are
sub-dominant contributions, from gravity-mediation, that are non-MFV
and that lead to potentially observable deviations in precision flavor
measurements.

In a previous work \cite{Hiller:2008sv}, we assumed that
the structure of the non-MFV terms is related to that of the Yukawa
couplings. Specifically, we assumed that there is an approximate
Froggatt-Nielsen (FN) symmetry \cite{Froggatt:1978nt} that leads to
selection rules which, in turn, dictate the structure of both the
Yukawa couplings and the soft supersymmetry breaking terms
\cite{Nir:1993mx,Leurer:1993gy}. In this work, in order to further
explore the questions formulated above, we take a different path,
where the structure of non-MFV terms is not related to the standard
model one. Specifically, we assume that the gravity-mediated
contributions to the squark masses-squared are anarchical, namely they
are all of the same order, with no special features.

Such unrelated structures might arise because the Yukawa couplings
come from the superpotential, while the soft masses-squared come from
the Kahler potential. The two sectors may have different dynamics, or
different selection rules from approximate symmetries. Model building
directions are suggested by the framework of Ref.
\cite{Aharony:2010ch} that involves a strongly coupled sector, which
is approximately conformally invariant and leads to large anomalous
dimensions of some of the quark fields over a large range of
energies. If the CFT sectors are not separable, the relevant Yukawa
couplings are suppressed (though not necessarily hierarchical), yet
the corresponding soft terms are anarchical. 

While we assume the anarchical structure for the quadratic squark
masses throughout this work, we investigate three different scenarios
for the trilinear scalar couplings (the $A$-terms): anarchical,
vanishing, or of a structure similar to the Yukawa couplings. We do so
because the $A$-terms, unlike the soft masses, come from the
superpotential, as do the Yukawa couplings, and furthermore they
transform under the flavor symmetry in precisely the same way as the
Yukawa couplings. It could thus well be that their structure is related
to the Yukawa sector, while the quadratic terms are not.

Our work here is aimed to answer the question formulated above, of
whether a relation between the new physics and the standard model
flavor structures exists at all. If the answer will end up being in
the affirmative, we will be able to go a step further. Indeed, one can
think of various mechanisms that would relate the two sectors, and the
final goal would be to distinguish between them and answer questions
such as the following:
\begin{itemize}
\item Does the flavor structure of the standard model come from an
  approximate symmetry, or from some dynamical mechanism? If it is a
  symmetry, is it Abelian or non-Abelian?
\end{itemize}
It should be interesting to pursue these questions in detail.

Within our framework, we impose the constraints that follow from low
energy flavor measurements, and obtain the upper bounds on possible
deviations that might still be discovered in the future. In this way,
we show how the future flavor measurements may provide answers to the
questions posed above, and by that lead to insights concerning the
underlying theory of supersymmetry breaking mediation (the new physics
flavor puzzle) and the theory of flavor (the standard model flavor
puzzle).

The outline of this paper is as follows. In Section \ref{sec:FCNCrev}
we set our notations for the supersymmetric flavor parameters and
review the FCNC and CP constraints. In Section \ref{sec:init-con} we
present the soft terms at the high and at the weak scale in our
framework of hybrid gauge-gravity mediation of supersymmetry breaking.
In Section \ref{sec:FNsum} we summarize the analysis of Ref.
\cite{Hiller:2008sv} of the implications of FCNCs on models in which
the gravity-mediated contributions are subject to an FN mechanism.
In this work we extend this study by considering $A$-terms as well.
Section \ref{sec:anarchy} contains the main bulk of our work. In this
Section we analyze the implications of FCNC and CP-violating processes
on models where gravity-mediated contributions are anarchical. In
Section \ref{sec:con} we discuss how, in the future, a pattern of
deviations from the standard model predictions for FCNCs can shed
light on basic flavor puzzles.  Further technical details are given in
Appendix \ref{sec:rge}, where we explain how the low
energy flavor-violating parameters are related to the high scale soft
supersymmetry breaking terms.

\section{FCNC and CP constraints on SUSY parameters}
\label{sec:FCNCrev}
Measurements of various low energy processes put strong indirect
restrictions on physics beyond the standard model. Here, we briefly
review the constraints from flavor-violating and CP-violating
processes on the SUSY parameters relevant to our analysis.

Supersymmetric models provide, in general, new sources of flavor
violation. These are most commonly analyzed in the basis in which the
corresponding (down or up) quark mass matrix and the neutral gaugino
vertices are diagonal. In this basis, which we label by a tilde, the
squark mass matrices are not necessarily flavor-diagonal, and have the
form
\beq \label{eq:delta-def}
\tilde q_{Mi}^*(\widetilde M_{\tilde q}^2)^{MN}_{ij}\tilde q_{Nj}=
(\tilde q_{Li}^*\ \tilde q_{Rk}^*)\left(\begin{array}{cc}
(\widetilde M^2_{\tilde q_L})_{ij} & \widetilde A^q_{il}v_q \cr
\widetilde A^q_{jk}v_q & (\widetilde M^2_{\tilde q_R})_{kl} \cr
  \end{array}\right) \left(\begin{array}{c}
    \tilde q_{Lj} \cr \tilde q_{Rl} \cr
  \end{array}\right) .
\eeq
Here, $M,N=L,R$ label the chirality, and $i,j,k,l=1,2,3$ are
generation indices. $\widetilde M^2_{\tilde q_L}$ and $\widetilde
M^2_{\tilde q_R}$ are the supersymmetry-breaking squark
masses-squared. The $\widetilde A^q$ parameters enter in the trilinear
scalar couplings $\widetilde A^q_{ij}\phi_q\widetilde q_{Li}\widetilde
q_{Rj}^*$, where $\phi_q$ $(q=u,d)$ is the $q$-type Higgs boson. The
latter develop a vacuum expectation value $v_q=\langle\phi_q\rangle$,
with $v=\sqrt{v_u^2 +v_d^2}\sim174$ GeV and $v_u/v_d=\tan \beta$.

In Eq.~(\ref{eq:delta-def}) we omit flavor-diagonal $F$- and $D$-term
contributions present in the minimal supersymmetric standard model
(MSSM) since they are not relevant to our analysis.  Note that the
$F$- and $D$-term contributions to the chirality-preserving mass terms
$(\widetilde M^2_{\tilde q_{L,R}})_{ii}$ are suppressed anyway by
$v^2/\tilde m_q^2$ with respect to the SUSY-breaking contributions,
where $\tilde m^2_q$ is a representative $q$-squark mass scale.

In the tilde-basis, flavor violation takes place through squark
mass insertions, bringing in factors of
\beq
(\delta_{ij}^q)_{MN}\equiv(\widetilde M^2_{\tilde
  q})^{MN}_{ij}/\tilde m_q^2 .
\eeq
It is useful to define also
\beq
\langle\delta^q_{ij}\rangle=\sqrt{(\delta_{ij}^{q})_{LL}
  (\delta^{q}_{ij})_{RR}}.
\eeq
The $\delta^q$ parameters cause flavor and, if complex, CP violation
beyond the standard model, and are constrained by indirect measurements.

In Table \ref{tab:exp} we compile the constraints on the
chirality-preserving $\delta^q_{MM}$ parameters obtained in Refs.
\cite{Masiero:2005ua,Gedalia:2009kh,Buchalla:2008jp}. Wherever
relevant, a mild phase suppression in the mixing amplitude is allowed,
namely we quote the stronger between the bounds on ${\cal
  R}e(\delta^q_{ij})$ and $3{\cal I}m(\delta^q_{ij})$.  The dependence
of these bounds on the average squark mass $m_{\tilde q}$, the ratio
$x\equiv m_{\tilde g}^2/m_{\tilde q}^2$ as well as the effect of
arbitrary CP violating phases can be found in Ref. \cite{Hiller:2008sv},
and references therein.  For the $D$ system, we use the recent
constraints of Ref. \cite{Gedalia:2009kh} incorporating updated CP-violating
effects.

\begin{table}[t]
\caption{The phenomenological upper bounds on the chirality-preserving
  couplings $(\delta_{ij}^{q})_{A}$ and
   on $\langle\delta^q_{ij}\rangle$, where $q=u,d$ and $A=LL,RR$.
   The constraints are given for $m_{\tilde q}=1$ TeV and $x\equiv m_{\tilde
   g}^2/m_{\tilde q}^2=1$. We assume that the phases could suppress the
   imaginary parts by a factor $\sim0.3$. The bound on
   $(\delta^{d}_{23})_{RR}$ is about 3 times weaker than that on
   $(\delta^{d}_{23})_{LL}$ (given in the table). The constraints on
   $(\delta^{d}_{12,13})_A$, $(\delta^{u}_{12})_A$ and $(\delta^{d}_{23})_A$
   are based on, respectively, Refs. \cite{Masiero:2005ua}, \cite{Gedalia:2009kh}
   and \cite{Buchalla:2008jp}.}
\label{tab:exp}
\begin{center}
\begin{tabular}{cc|cc} \hline\hline
\rule{0pt}{1.2em}%
$q$\ & $ij\ $\ &  $(\delta^{q}_{ij})_A$ & $\langle\delta^q_{ij}\rangle$ \cr \hline
$d$ & $12$\ & $\ 0.03\ $ & $\ 0.002\ $ \cr
$d$ & $13$\ & $\ 0.2\ $ & $\ 0.07\ $ \cr
$d$ & $23$\ & $\ 0.6\ $ & $\ 0.2\ $ \cr
$u$ & $12$\ & $\ 0.1\ $ & $\ 0.008\ $ \cr
\hline\hline
\end{tabular}
\end{center}
\end{table}

For large $\tan\beta$, additional constraints with respect to those in
Table \ref{tab:exp} arise.  In particular the effects of neutral Higgs
exchange in $B_s$ and $B_d$ mixing are important.  For instance, for $\tan
\beta =30$ and $x=1$ \cite{Isidori:2002qe,Hiller:2008sv}
\beq \label{eq:bmixbounds}
\langle \delta^d_{13}\rangle  < 0.01 \cdot \left( \frac{M_{A^0}}{200
    \, \mbox{GeV}} \right) , ~~~~~
\langle \delta^d_{23} \rangle < 0.04 \cdot \left( \frac{M_{A^0}}{200
    \, \mbox{GeV}} \right) ,
\eeq
where $M_{A^0}$ denotes the pseudoscalar Higgs mass, and the above
bounds scale roughly as $(30/\tan \beta)^2$.  A more detailed
discussion including chargino contributions and the impact of rare
decays can be found in Ref. \cite{Hiller:2008sv}.

The experimental constraints on the $(\delta^q_{ij})_{LR}$ parameters
in the quark-squark sector are presented in Table \ref{tab:expLRme}.
Very strong constraints apply for the imaginary part of $(\delta^q_{11})_{LR}$
from electric dipole moments (EDMs). The bounds given here correspond to
the experimental upper limit on the EDM of the neutron, $d_n < 2.9
\cdot 10^{-26}\ e\ {\rm cm}$ \cite{Amsler:2008zzb}.  For $x=4$ and a
phase smaller than 0.1, the EDM constraints on
$(\delta^{u,d}_{11})_{LR}$ are weakened by a factor of $\sim 6$.

\begin{table}[t]
\caption{The phenomenological upper bounds on the chirality-mixing parameters
$(\delta_{ij}^{q})_{NM}$, $N\neq M$ and $q=u,d$.
   The constraints are given for $m_{\tilde q}=1$ TeV and $x\equiv m_{\tilde
   g}^2/m_{\tilde q}^2=1$.  We assume that the phases could suppress the
   imaginary parts by a factor $\sim0.3$. The constraints on
   $(\delta^{d}_{12,13})_{NM}$, $(\delta^d_{23})_{NM}$ and $(\delta^{u}_{12})_{NM}$, and
   $(\delta^{q}_{ii})_{NM}$
   are based on, respectively, Refs.~\cite{Masiero:2005ua},
    \cite{Buchalla:2008jp} and
   \cite{Gabbiani:1996hi,Amsler:2008zzb}. The bounds are the same for
   $\delta^q_{LR}$ and $\delta^q_{RL}$, except for $(\delta^d_{12})_{NM}$,
   where the bound in parentheses refers to $NM=RL$.}
\label{tab:expLRme}
\begin{center}
\begin{tabular}{cc|c} \hline\hline
\rule{0pt}{1.2em}%
$q$\ & $ij$\ & $(\delta^{q}_{ij})_{LR ~ (RL)} $\cr \hline
$d$ & $12$\  &    $\ 2 \cdot 10^{-4} \, (0.002) \ $  \cr
$d$ & $13$\  &  $\ 0.08 \ $  \cr
$d$ & $23$\  &  $\ 0.01 \ $ \cr
$d$ & $11$\  & $4.7\cdot10^{-6}$  \cr
$u$ & $11$\  &  $9.3\cdot 10^{-6}$  \cr
$u$ & $12$\  &  $\ 0.02 \ $ \cr
\hline\hline
\end{tabular}
\end{center}
\end{table}

\section{Hybrid Gauge-Gravity Models}
\label{sec:init-con}
We consider supersymmetric models with gauge-mediated SUSY breaking in
the presence of contributions induced by gravity at the Planck scale.
While the former follows the flavor structure dictated by the one
already present in the standard model, the latter allows, in general,
for further intergenerational sfermion flavor mixings.

In Section \ref{sec:hiscale} we set our initial conditions at the
scale of gauge-mediation, $m_M$, and in Section \ref{ssec:delparam}
give approximate analytical expressions for the flavor and
CP-violating $\delta^q$ parameters defined in Section
\ref{sec:FCNCrev}.  The soft terms at $m_M$ and at the electroweak
scale, $m_Z$, are linked by the MSSM renormalization group (RG)
equations \cite{Martin:1993zk}.  Details on the RG evolution (RGE) are
given in Appendix \ref{sec:rge}.

\subsection{High scale \label{sec:hiscale}}
In the discussed hybrid setup, the soft terms at the scale of
gauge-mediation can be written as
\beqa\label{eq:inicon}
M^2_{\tilde Q_L}(m_M)&=&\tilde m^2 ({\bf 1}+ r X_{q_L}),\no\\
M^2_{\tilde D_R}(m_M)&=&\tilde m^2 ({\bf 1}+r X_{d_R}),\no\\
M^2_{\tilde U_R}(m_M)&=&\tilde m^2 ( {\bf 1}+r X_{u_R}).
\eeqa
Here, $\tilde m$ is the typical scale of the gauge-mediated
contribution to the soft terms, which is universal in the limit of
neglecting $\alpha_{1,2}/\alpha_3$ effects, where $\alpha_i =g_i^2/(4
\pi)$, and $g_{1,2} \, (g_3)$ denote the gauge couplings of the
electroweak sector (strong interaction).  Above, the coefficient
$r\lsim1$ parameterizes the ratio between the gravity-mediated and the
gauge-mediated contributions.  Gravity-mediation induces also
trilinear terms of the form \beqa\label{eq:iniconA}
A^u(m_M)&=& \tilde m\sqrt{r}\ Z_{A_u},\no\\
A^d(m_M)&=& \tilde m\sqrt{r}\ Z_{A_d} .  \eeqa

While the gauge-mediated initial conditions are flavor-universal,
generically the $X_{q_A}$ and $Z_{A_q}$ matrices carry non-trivial
flavor structure. It is the goal of this paper to explore the
different phenomenology following from different ans\"atze for the
generational structure of the $X_{q_A}, Z_{A_q}$ matrices.

\subsection{The $\delta^q$ parameters}
\label{ssec:delparam}
The initial conditions Eqs.~(\ref{eq:inicon}) and (\ref{eq:iniconA})
hold at the high scale $m_M$, while flavor-changing and CP-violating
processes restrict the weak scale parameters $(\delta^q_{ij})_{NM}$.
Thus, the effect of the RG evolution on the soft terms must be
evaluated.  Furthermore, the $(\delta^q_{ij})_{NM}$ parameters are
read off from the low energy soft terms in the basis in which the
quark mass matrices and gluino couplings are diagonal, which differs
from the flavor basis.  The requisite rotation of the squarks leaves
the parametric pattern of the $X_{ij}$ and $Z_{ij}$ unchanged.

In Appendix \ref{sec:rge}, we give a detailed derivation, given our
framework and various approximations, of the low energy flavor
parameters. It leads to the following expressions at the weak scale:
\beqa\label{eq:delgen}
(\delta^{u}_{12})_{LL} & \sim & \frac{1}{r_3}
{\rm max}\{r (X_{u_L}+Z_{A_u} Z_{A_u}^\dagger
+ Z_{A_d} Z_{A_d}^\dagger)_{12},   c_d y_b^2|V_{ub} V_{cb}^*|\} ,\no \\
(\delta^{d}_{12})_{LL} & \sim & \frac{1}{r_3}
{\rm max}\{r (X_{d_L}+Z_{A_u} Z_{A_u}^\dagger
+ Z_{A_d} Z_{A_d}^\dagger)_{12}, c_u y_t^2|V_{ts} V_{td}^*|\} ,\no \\
(\delta^{u}_{i3})_{LL}& \sim & \frac{1}{r_3}
{\rm max} \{r(X_{u_L}+Z_{A_u} Z_{A_u}^\dagger
+ Z_{A_d} Z_{A_d}^\dagger)_{i3}, c_d y_b^2 |V_{ib}V_{tb}^*|\} ,\no\\
(\delta^{d}_{i3})_{LL}& \sim & \frac{1}{r_3}
{\rm max} \{r(X_{d_L}+Z_{A_u} Z_{A_u}^\dagger
+ Z_{A_d} Z_{A_d}^\dagger)_{i3}, c_u y_t^2|V_{tb}V_{ti}^*|\},\no\\
(\delta^q_{ij})_{RR}& \sim &\frac{r}{r_3}  (X_{q_R}+ Z_{A_q}^\dagger
Z_{A_q})_{ij} , ~( i \neq j),\no \\
(\delta^u_{ij})_{LR}& \sim &
(Z_{A_u})_{ij} \sqrt r v\sin\beta/(r_3\tilde m),  \no \\
(\delta^d_{ij})_{LR}& \sim &
(Z_{A_d})_{ij} \sqrt r v\cos \beta/(r_3\tilde m),
\eeqa
where $V_{ij}$ are CKM elements, $y_t,y_b$ denotes the top, bottom
Yukawa, respectively, $i=1,2$ and $j=1,2,3$. The factor $r_3$ captures
the effect of RGE corrections to the diagonal elements of the soft
squark mass matrices $(M^2_{\tilde q_A})_{ii}$ and is defined in Eq.
(\ref{defrthree}).  Numerically, $r_3={\cal{O}}(1-10)$, depending on
the initial conditions, the scale of SUSY breaking and hidden sector
effects. In minimal models, typically $r_3 \sim 3$. The coefficients
$c_u,c_d$ can be of ${\cal{O}}(1)$ for $m_M$ near the GUT scale and
are all negative.  The expressions Eq.~(\ref{eq:delgen}) hold also for
$(\delta^q_{jj})_{LR}$ up to MSSM $F$-term contributions. Throughout
this work, the ``$\sim$'' sign implies a similar parametric suppression
but with generally different ${\cal O}(1)$ complex coefficients.

\section{FN symmetry in the gravity sector}
\label{sec:FNsum}
A mediation mechanism allowing non-MFV contributions to the soft SUSY
breaking terms in which flavor-changing terms are nonetheless
suppressed was considered in Ref. \cite{Feng:2007ke}.  In such a
setup, the gauge-mediation contributions are dominant, but
gravity-mediation contributions are non-negligible. In Ref.
\cite{Feng:2007ke}, the structure of the gravity-mediated
contributions was not arbitrary, but rather set by the same
approximate horizontal symmetry which explains the smallness of the
Yukawa couplings {\it \`a la} Froggatt-Nielsen
\cite{Froggatt:1978nt,Leurer:1993gy}.

In Section \ref{ssec:FCNCsumFN} we summarize the implications of FCNC
constraints in such hybrid FN models as obtained in Ref.
\cite{Hiller:2008sv}, updating these to include the more recent
constraints in the $D$ system. In Section \ref{sec:A-FN} we
investigate the soft breaking $A$-terms in the presence of the flavor
symmetry.

\subsection{Flavor breaking in hybrid FN models}
\label{ssec:FCNCsumFN}
We now summarize the results of Ref. \cite{Hiller:2008sv}, in which
the gravity-mediated contributions to the soft supersymmetry breaking
terms are assumed to be subject to the selection rules of the FN
symmetry.  Within the simplest FN models, with a single horizontal
$U(1)_H$, the parametric structure of the gravity-mediated
contributions to the soft terms (\ref{eq:inicon}) is given by
\beq\label{eq:fnxij}
(X_{q_{L,R}})_{ii}\sim1,\ \ \ (X_{q_L})_{ij}\sim|V_{ij}|,\ \ \
(X_{q_R})_{ij}\sim\frac{m_{q_i}/m_{q_j}}{|V_{ij}|}\ \ \ (i<j), ~~q=u,d,
\eeq
where $m_{q_i}$ denotes the $i$th generation $q$-type quark mass.
For the non-MFV contributions, in which there are uncertainties of
order one, we use, for example, $V_{13}$ to represent a parametric
suppression that is similar to that of $V_{ub}$ or $V_{td}$.  For the
MFV contributions, we use notations such as $V_{td}$ to denote the
actual contributing CKM element.

Imposing the flavor structure of Eq.~(\ref{eq:fnxij}) on the
expressions (\ref{eq:delgen}), we obtain the order of magnitude
estimates for the $\delta^q_{ij}$ parameters presented in Table
\ref{tab:the}. The $\hat r$ parameter is defined as
\beq\label{eq:rhat}
\hat r\equiv{\rm max}\{r,y_b^2\}.
\eeq

\begin{table}[t]
\caption{The order of magnitude estimates for
  $(\delta_{ij}^{d,u})_{LL,RR}$ and $\langle\delta^{d,u}_{ij}\rangle$ in
  the hybrid gauge-gravity models with FN structure
  \cite{Hiller:2008sv}. The numerical estimates are
  obtained using quark masses at the scale $m_Z$ \cite{Xing:2007fb},
  and taking $r_3=3$. All results scale as $(3/r_3)$.  For
  $\langle\delta^d_{i3}\rangle$ we use $|V_{i3}| \sim |V_{ti}|$.}
\label{tab:the}
\begin{center}
\begin{tabular}{ll|ccc} \hline\hline
\rule{0pt}{1.2em}%
$q$ & $ij$\ &  $(\delta^q_{ij})_{LL}$ &  $(\delta^q_{ij})_{RR}$ &
$\langle\delta^q_{ij}\rangle$ \cr \hline
$d$ & $12$\ & $(r/r_3)|V_{12}|\sim0.08r$ &
$\frac{(r/r_3)(m_d/m_s)}{|V_{12}|}\sim0.08r$
& $(r/r_3)\sqrt{m_d/m_s}\sim0.08r$ \cr
$d$ & $13$\ & $y_t^2 |V_{td}^* V_{tb}|/r_3\sim0.003$ &
$\frac{(r/r_3)(m_d/m_b)}{|V_{13}|}\sim0.08 r$
& $y_t \sqrt{r m_d/m_b}/r_3\sim0.01\sqrt{r}$ \cr
$d$ & $23$\ & $y_t^2 |V_{ts}^* V_{tb}|/r_3\sim0.01$ &
$\frac{( r/r_3)(m_s/m_b)}{|V_{23}|}\sim0.2 r$
& $y_t \sqrt{r m_s/m_b}/r_3\sim0.05\sqrt{ r}$ \cr
$u$ & $12$\ & $(r/r_3)|V_{12}|\sim0.08r$ &
$\frac{(r/r_3)(m_u/m_c)}{|V_{12}|}\sim0.003r$
& $(r/r_3)\sqrt{m_u/m_c}\sim0.02r$  \cr
$u$ & $13$\ & $(\hat r/r_3)|V_{13}|\sim0.001 \hat r$ &
$\frac{(r/r_3)(m_u/m_t)}{|V_{13}|}\sim0.0006 r$
& $\sqrt{r \hat r m_u/m_t}/r_3\sim0.0009\sqrt{ r \hat r}$ \cr
$u$ & $23$\ & $(\hat r/r_3)|V_{23}|\sim0.01 \hat r$ &
$\frac{(r/r_3)(m_c/m_t)}{|V_{23}|}\sim0.03 r$
& $\sqrt{r \hat r m_c/m_t}/r_3\sim0.02\sqrt{ r \hat r}$ \cr
\hline\hline
\end{tabular}
\end{center}
\end{table}

Comparing the phenomenological constraints of Table \ref{tab:exp} to
the theoretical predictions of Table \ref{tab:the}, we obtain upper
bounds on $r$ and on $\hat r$. The strongest bound on $r$ comes from
the $\langle\delta^d_{12}\rangle$ parameter, {\it i.e.} from the
neutral Kaon system:
\beq \label{eq:FNroverr3}
r/r_3\lsim0.01-0.03.
\eeq
Here we use $m_{\tilde q}=1$ TeV; for lighter $m_{\tilde q}$ the
bounds would be stronger by $m_{\tilde q}/(1$ TeV). The stronger bound
corresponds to $x=1$ and a phase of order $0.3$, while the weaker
bound corresponds to $x=4$ and a phase smaller than $0.1$.  Since the
$\hat r$ parameter affects only the $\delta^u_{i3}$ parameters, there
is no phenomenological constraint on its size, and it is only bounded
by its definition:
\beq\label{eq:FNrhatoverr3}
r\leq \hat r \lsim 1.
\eeq
For small values of $\tan \beta$, $\hat r =r$ and
Eq.~(\ref{eq:FNroverr3}) applies to $\hat r$.  Inserting
$r/r_3\lsim0.03$ and $r\leq\hat r\lsim1$ into the predictions of Table
\ref{tab:the}, we obtain the upper bounds on the $\delta^q_{ij}$
presented in Table \ref{tab:upb}.

\begin{table}[t]
\caption{The order of magnitude upper bounds on
   $(\delta_{ij}^{d,u})_{LL,RR}$ and
   $\langle\delta^{d,u}_{ij}\rangle$ for $r/r_3\lsim0.03$ in hybrid FN
   models \cite{Hiller:2008sv}.
   Entries in parentheses are independent of
   $r$, therefore representing estimates rather than upper bounds, and
   scale as $(3/r_3)$. The bounds on $\langle \delta^d_{13,23}\rangle$
   scale as $\sqrt{3/r_3}$. The bounds on $(\delta^u_{i3})_{LL}$
   [$\langle\delta^u_{i3}\rangle$] correspond to $\hat r\sim1$ and
   scale as $(3/r_3)$ [$\sqrt{3/r_3}$]; if $\hat r=r$, these bounds
    are a factor of 10 [$\sqrt{10}$] stronger and do not scale
   with $r_3$.     }
\label{tab:upb}
\begin{center}
\begin{tabular}{ll|ccc} \hline\hline
\rule{0pt}{1.2em}%
$q$ & $ij$\ &  $(\delta^q_{ij})_{LL}$ &  $(\delta^q_{ij})_{RR}$ &
$\langle\delta^q_{ij}\rangle$ \cr \hline
$d$ & $12$\ & $0.007$ & $0.007$ & $0.007$ \cr
$d$ & $13$\ & $[0.003]$ & $0.007$ & $0.003$ \cr
$d$ & $23$\ & $[0.01]$ & $0.01$ & $0.01$ \cr
$u$ & $12$\ & $0.007$ & $0.0003$ & $0.001$  \cr
$u$ & $13$\ & $0.001$ & $0.00005$ & $0.0003$ \cr
$u$ & $23$\ & $0.01$ & $0.003$ & $0.006$ \cr
\hline\hline
\end{tabular}
\end{center}
\end{table}

The maximal possible effects in the neutral $B_d,\ B_s$ and $D$
systems are thus as follows (for $r_3=3$):
\beq\label{eq:FN-reach}
\begin{array}{lc}
B_d: & |M_{12}^{\rm susy}/M_{12}^{\rm exp}|\lsim0.002,\\
B_s: & |M_{12}^{\rm susy}/M_{12}^{\rm exp}|\lsim0.005,\\
D: & |M_{12}^{\rm susy}/M_{12}^{\rm exp}|\lsim0.03. \end{array}
\eeq
The sensitivity in the $D$ system is slightly modified in comparison
to Ref. \cite{Hiller:2008sv} due to the use of the updated analysis of
Ref. \cite{Gedalia:2009kh}, see also Table \ref{tab:exp}.

The mixing amplitudes of the $B_{d,s}$ mesons can be significantly
enhanced for low $M_{A^0}$ and large $\tan\beta$. By comparing the
phenomenological constraints of Eq. (\ref{eq:bmixbounds}) to the
predictions of Table \ref{tab:the} one finds (for $r_3=3$, $\tan
\beta=30$ and $M_{A^0}=200 \, \mbox{GeV}$):
\beq\label{eq:FN-reach-largetb}
\begin{array}{lc}
B_d: & |M_{12}^{\rm susy}/M_{12}^{\rm exp}|\lsim0.10,\\
B_s: & |M_{12}^{\rm susy}/M_{12}^{\rm exp}|\lsim0.13.
\end{array}
\eeq
The effects on these systems are maximized when the RGE suppression is
minimal. Further details, such as a variant of FN models with
holomorphic zeros, where the gravity-mediated contribution to the
$D^0-\overline D^0$ mixing amplitude can be of ${\cal O}(1)$, can be
found in Ref. \cite{Hiller:2008sv}.

\subsection{$A$-terms in hybrid FN models \label{sec:A-FN}}
Going beyond Ref. \cite{Hiller:2008sv}, we investigate here the
trilinear $A$-terms in hybrid models with a FN flavor symmetry. In
such scenarios, the $A$-terms follow the same parametric suppression
as the corresponding Yukawa matrices $Y$. At the high, messenger
scale:
\beq\label{eq:yuklik}
(A^{u,d})_{ij}(m_M)\sim \sqrt{r}\ \tilde{m}Y^{u,d}_{ij}.
\eeq
Since the $A$-terms are only similar in texture to the Yukawa
matrices, but not proportional to them, rotating to the mass basis
leaves the $A$-terms undiagonalized ($q=u,d$)
\beqa\label{eq:ApreyukFN}
(Z_{A_q})_{ij}\sim Y_{ij}^{q}\sim V_{ij}{m_{q_j}}/{v_q}.
\eeqa
The resulting chirality-mixing $\delta^q_{LR}$ parameters, see
Eq.~(\ref{eq:delgen}), are less important for flavor physics than the
chirality-preserving $\delta^q_{LL,RR}$ parameters. If CP-violating,
the $\delta^q_{LR}$ induce a neutron EDM allowed by
Eq.~(\ref{eq:FNroverr3}):
\beq \label{eq:FNEDM}
|d_{n}^{\rm susy}/d_{n}^{\rm exp}|\lsim 0.02 \, (0.002) ,
\eeq
where the first value corresponds to $x=1$ and a phase suppression in
$(\delta^q_{11})_{LR}$ of $\sim 0.3$, and the value in parentheses is
obtained for $x=4$ and a phase suppression of $\sim 0.1$.

\section{Anarchy in the gravity sector}
\label{sec:anarchy}
Thus far, we have considered flavor-changing processes within
supersymmetric models with hybrid gauge-gravity mediation, in which
the structure of the gravity contributions is dictated by the FN
mechanism.  However, the gravity sector need not obey such selection
rules and may, for example, be of anarchical character.  By anarchy we
mean structure-less gravity contributions, such that all terms of the
(hermitian) matrices in Eq.~(\ref{eq:inicon}) obey
\beq \label{eq:X-ana}
(X_{q_A})_{ij} \sim {\cal O}(1),
\eeq
and carry, in general, order one CP-violating phases. In particular we
do not consider accidental suppressions in the magnitude of individual
matrix elements. We now study which measurements can reveal the
existence of such anarchical models.

Assuming anarchical structure for the squark masses-squared, one can
still consider various structures for the trilinear scalar couplings.
The effect of non-vanishing $A$-terms is two-fold: First, the RG
evolution of the soft terms is modified, and second, chirality-mixing
processes may get direct contributions from these terms. We explore
three different scenarios for the $A$-terms:
\begin{enumerate}
\item Section \ref{ssec:noA}: vanishing $A$-terms;
\item Section \ref{ssec:anA}: anarchical $A$-terms;
\item Section \ref{ssec:yukA}: Yukawa-like textured $A$-terms.
\end{enumerate}

Before we start a detailed discussion, a comment regarding the MFV
terms is in order. In the current context of an anarchical texture in
the $X_{q_A}$ matrices Eq.~(\ref{eq:X-ana}), non-MFV effects are
non-negligible in the $\delta^q_{LL}$ parameters provided $r \gsim
y_t^2 \lambda^5\sim 3 \cdot 10^{-4}$, where $\lambda\sim|V_{12}| \sim
0.2$, as can be seen from Eq.~(\ref{eq:delgen}).  This is a weaker
condition than in the analogous FN case, where interesting,
gravity-dominated effects require $r \gsim y_t^2 \lambda^4 \sim 2
\cdot 10^{-3}$.  This in turns implies that for the gravity-mediated
contributions to have observable consequences, the messenger scale can
in principle be lower in the anarchical setup than in the framework
with a FN flavor structure.

\subsection{Vanishing $A$-terms}
\label{ssec:noA}
We consider anarchical models with vanishing $A$-terms at the
messenger scale:
\beq \label{eq:Azero}
A^{u,d}(m_M)=0.
\eeq
Hence, our starting
point is Eq.~(\ref{eq:delgen})  with
\beq\label{eq:xzvani}
(X_{q_A})_{ij}={\cal O}(1),\ \ \ Z_{A_q}=0.
\eeq
We obtain the following flavor parameters, at the
$m_Z$-scale ($i=1,2$):
\beqa\label{eq:delql}
\begin{split}
(\delta^{u}_{12})_{LL} & \sim   \frac{r}{r_3} , \ \ \
(\delta^{d}_{12})_{LL}  \sim   \frac{r}{r_3} , \\
(\delta^{u}_{i3})_{LL} & \sim    \frac{\hat r_i}{r_3} , \ \ \
(\delta^{d}_{i3})_{LL}  \sim    \frac{\hat r^u_i}{r_3} , \\
\end{split}
\eeqa
and ($i\neq j,\ j=1,2,3$):
\beqa\label{eq:delqr}
(\delta^q_{ij})_{RR} & \sim&  \frac{r}{r_3},
\eeqa
where
\beq\label{eq:rhat}
\hat r_i\equiv{\rm max}\{r,y_b^2V_{ib} V_{tb}^*\},\ \ \
\hat r^u_i\equiv{\rm max}\{r,y_t^2V_{ti}^* V_{tb}\}.
\eeq
The resulting order of magnitude estimates for the
chirality-preserving $\delta^q_{ij}$ parameters are presented in Table
\ref{tab:andel}. The chirality-mixing $(\delta^q_{ij})_{LR}$
parameters play no role in constraining SUSY flavor in this scenario.

\begin{table}[t]
\caption{The order of magnitude estimates for
  $(\delta_{ij}^{d,u})_{A}$, $A=LL,RR$ and $\langle\delta^{d,u}_{ij}\rangle$ in
  the models defined by Eq. (\ref{eq:xzvani}). The numerical estimates are
  obtained using quark masses at the scale $m_Z$ \cite{Xing:2007fb}
  and for $r_3=3$. All numerical estimates scale as $(3/r_3)$. }
\label{tab:andel}
\begin{center}
\begin{tabular}{ll|ccc} \hline\hline
\rule{0pt}{1.2em}%
$q$ & $ij$\ &  $(\delta^q_{ij})_{LL}$ &  $(\delta^q_{ij})_{RR}$ &
$\langle\delta^q_{ij}\rangle$ \cr \hline
$d$ & $12$\ & $r/r_3\sim0.3r$ & $r/r_3\sim0.3r$
& $r/r_3\sim0.3r$ \cr
$d$ & $13$\ & $\hat r^u_1/r_3\sim 0.3\hat r^u_1$ &
$r/r_3\sim0.3 r$
& $\sqrt{\hat r^u_1 r}/r_3\sim0.3\sqrt{\hat r^u_1 r}$ \cr
$d$ & $23$\ & $\hat r^u_2/r_3\sim0.3\hat r^u_2$ &
$r/r_3\sim0.3 r$
& $\sqrt{\hat r^u_2 r}/r_3\sim0.3\sqrt{\hat r^u_2 r}$ \cr
$u$ & $12$\ & $r/r_3\sim0.3r$ & $r/r_3\sim0.3r$
& $r/r_3\sim0.3r$  \cr
$u$ & $13$\ & $\hat r_1/r_3\sim0.3\hat r_1$ & $r/r_3\sim0.3r$
& $\sqrt{\hat r_1 r}/r_3\sim0.3\sqrt{\hat r_1 r}$ \cr
$u$ & $23$\ & $\hat r_2/r_3\sim0.3\hat r_2$ & $r/r_3\sim0.3 r$
& $\sqrt{\hat r_2 r}/r_3\sim0.3\sqrt{\hat r_2 r}$ \cr
\hline\hline
\end{tabular}
\end{center}
\end{table}

By comparing the phenomenological constraints of Table \ref{tab:exp}
to the order of magnitude predictions of Table \ref{tab:andel}, we
obtain an upper bound on $r$ and find the maximal possible effects in
this scenario.  The strongest bound on $r$ comes from the neutral Kaon
system constraint on $\langle \delta^d_{12} \rangle$:
\beq\label{eq:rbound}
r/r_3\lsim 0.002-0.006.
\eeq
The range corresponds to the same assumptions that enter
Eq.~(\ref{eq:FNroverr3}).

The $\hat r_i$ ($\hat r^u_i$ ) parameters affect only the
$\delta^u_{i3}$ ($\delta^d_{i3}$) parameters, and they are bounded
only be their definition:
\beqa\label{eq:rhatb}
\hat r_1& ={\rm\max}\{r, y_b^2 V_{ub} V_{tb}^*\}
\sim {\rm max}\{r,0.004y_b^2\},\nonumber \\
\hat r_2& ={\rm max}\{r,y_b^2 V_{cb} V_{tb}^*\} \sim {\rm
  max}\{r,0.04y_b^2\},
\eeqa
and
\beqa\label{eq:rhatdb}
\hat r^u_1& ={\rm\max}\{r, y_t^2  V_{td}^*  V_{tb} \} \sim {\rm max}\{r,0.009\},
\nonumber \\
\hat r^u_2& ={\rm max}\{r,y_t^2  V_{ts}^*  V_{tb} \} \sim {\rm max}\{r,0.04\}.
\eeqa
Inserting the bounds (\ref{eq:rbound}), (\ref{eq:rhatb}) and
(\ref{eq:rhatdb}) into the predictions of Table \ref{tab:andel}, we
obtain the upper bounds on the $\delta^q_{ij}$ given in Table
\ref{tab:anupb}.

\begin{table}[t]
\caption{The order of magnitude upper bounds on
   $(\delta_{ij}^{d,u})_{LL,RR}$ and
   $\langle\delta^{d,u}_{ij}\rangle$ corresponding to $r/r_3\lsim0.006$ and
   the bounds of Eqs.~(\ref{eq:rhatb},\ref{eq:rhatdb}). Entries with
   an $r_3$ dependence are indicated so.}
\label{tab:anupb}
\begin{center}
\begin{tabular}{ll|ccc} \hline\hline
\rule{0pt}{1.2em}%
$q$ & $ij$\ &  $(\delta^q_{ij})_{LL}$ &  $(\delta^q_{ij})_{RR}$ &
$\langle\delta^q_{ij}\rangle$ \cr \hline
$d$ & $12$\ & $0.006$ & $0.006$ & $0.006$ \cr
$d$ & $13$\ & ${\rm max}\{0.006,0.003(3/r_3)\}$ & $0.006$ & ${\rm
  max}\{0.006,0.004\sqrt{3/r_3}\}$ \cr
$d$ & $23$\ & ${\rm max}\{0.006,0.01(3/r_3)\}$ & $0.006$ & ${\rm
  max}\{0.006,0.009\sqrt{3/r_3}\}$ \cr
$u$ & $12$\ & $0.006$ & $0.006$ & $0.006$  \cr
$u$ & $13$\ & ${\rm max}\{0.006,0.001 y_b^2(3/r_3)\}$ & $0.006$ &
${\rm max}\{0.006,0.003 y_b\sqrt{3/r_3}\}$ \cr
$u$ & $23$\ & ${\rm max}\{0.006,0.01y_b^2(3/r_3)\}$ & $0.006$ & ${\rm
  max}\{0.006,0.009 y_b\sqrt{3/r_3}\}$ \cr
\hline\hline
\end{tabular}
\end{center}
\end{table}

We learn that the maximal possible effects in the neutral $B_d$,
$B_s$ and $D$ systems are as follows (for $r_3=3$):
\beq\label{eq:anarchy-reach}
\begin{array}{lc}
B_d: & |M_{12}^{\rm susy}/M_{12}^{\rm exp}|\lsim0.007,\\
B_s: & |M_{12}^{\rm susy}/M_{12}^{\rm exp}|\lsim0.002,\\
D: & |M_{12}^{\rm susy}/M_{12}^{\rm exp}|\lsim0.6. \end{array}
\eeq
Note that these systems do not depend on $\hat r_i$, and so are
independent of $y_b$.

We emphasize the following points:
\begin{itemize}
\item The bound in the $D$ system comes from $\langle
  \delta^u_{12}\rangle$ and is $r_3$ independent.
\item For $r_3={\cal O}(1-10)$, the bound in the $B_s$ system comes
  from $\langle\delta^d_{23}\rangle$.  For $r_3={\cal O}(1-7)$ it
  scales with $3/r_3$; for $r_3>7$ it does not scale with $r_3$.
\item For $r_3={\cal O}(1-10)$, the bound in the $B_d$ system comes
  from $\langle\delta^d_{13}\rangle$.  For $r_3={\cal O}(1-1.5)$ it
  scales with $3/r_3$; for $r_3>1.5$ it does not scale with $r_3$.
\end{itemize}

\begin{figure}
\includegraphics[scale=1.0]{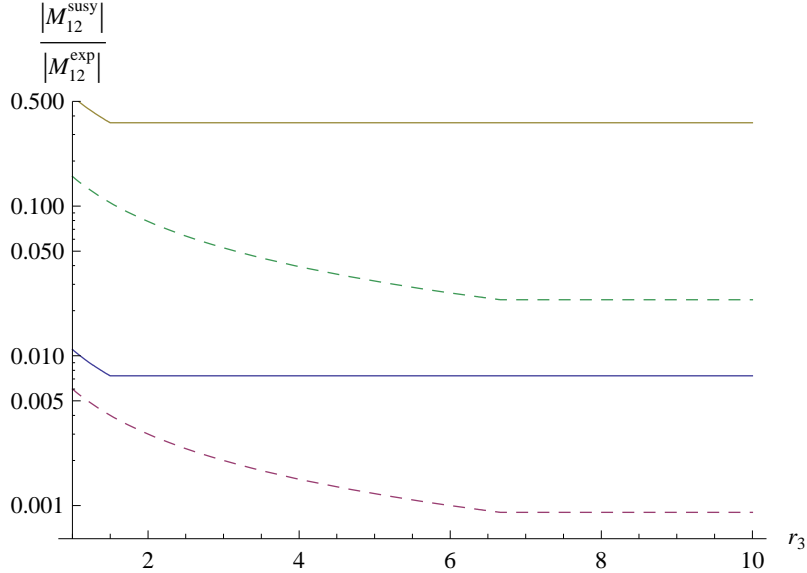}
\caption{\label{fig:Bboundall} Maximum reach in $B_d$ (solid) and
$B_s$ (dashed) mixing, $|M_{12}^{\rm susy}/M_{12}^{\rm exp}|$,
as a function of the RGE-factor $r_3$, for hybrid anarchy models with
vanishing soft trilinear couplings as in Eq.~(\ref{eq:Azero}). The two
upper curves correspond to $\tan \beta=30$ and
$M_{A^0}=200 \, \mbox{GeV}$, while the two lower ones correspond to
low $\tan\beta$.}
\end{figure}

As mentioned in Section \ref{sec:FCNCrev}, for large $\tan \beta$ and
low $M_{A^0}$, the $B_{d,s}$ mixing amplitudes can be significantly
enhanced. Indeed, comparing the constraints of
Eq.~(\ref{eq:bmixbounds}) to Table \ref{tab:andel}, we obtain for
$r_3=3$ (and $\tan \beta=30$, $M_{A^0}=200 \, \mbox{GeV}$):
\beq\label{eq:FN-reach-largetb}
\begin{array}{lc}
B_d: & |M_{12}^{\rm susy}/M_{12}^{\rm exp}|\lsim0.36,\\
B_s: & |M_{12}^{\rm susy}/M_{12}^{\rm exp}|\lsim0.05 .
\end{array}
\eeq
The $r_3$ dependence of upper bounds on the supersymmetric
contributions to $B_d$ and $B_s$ mixings is shown in Fig.~\ref{fig:Bboundall}.

\subsection{Anarchical $A$-terms}\label{ssec:anA}
We consider scenarios with anarchical $A$-terms at the messenger
scale:
\beq\label{eq:A}
(A^{u,d})_{ij}(m_M)\sim \sqrt{r}\ \tilde{m},
\eeq
so that
\beq\label{eq:xzanar}
(X_{q_A})_{ij}={\cal O}(1),\ \ \ (Z_{A_q})_{ij}={\cal O}(1).
\eeq

We insert the anarchical structure (\ref{eq:xzanar}) into the
expressions of the $\delta^q$ parameters in Eq.~(\ref{eq:delgen}). We
then compare these predictions to the bounds given in the Tables
\ref{tab:exp} and \ref{tab:expLRme}.  Note that these bounds are
obtained for $m_{\tilde{q}}=1$ TeV. Recalling that parametrically
$m_{\tilde q}^2\sim r_3 \tilde m^2$, we use $\tilde m\sim 1\ {\rm
  TeV}/\sqrt{r_3}$.

The anarchical $A$-terms induce ${\cal O}(1)$ changes in the $X_{q_A}$
matrices in the chirality preserving LL and RR blocks of the squark
mass-squared matrices.  Hence, the order of magnitude estimates for
the chirality-preserving $(\delta^q_{ij})_{LL,RR}$ parameters given in
Eqs.~\eqref{eq:delql} and \eqref{eq:delqr} remain standing.  The new
ingredients in this analysis with full anarchy are the
chirality-mixing parameters $\delta^q_{LR}$.

We find that the strongest bound on $r/r_3$ comes from the EDM
constraints on $(\delta^{u,d}_{11})_{LR}$:
\beq\label{eq:LRboundEDM}
\begin{split}
&r/r_3<\left\{\begin{array}{ll}
(7.3\cdot 10^{-10}-2.8 \cdot 10^{-8}) \,
\frac{1+\tan^2\beta}{(Z_{A_d})_{11}^2}
& \ \ {\rm down\ sector},\\
(2.9\cdot 10^{-9}-1.1\cdot 10^{-7})\,
\frac{1+\tan^2\beta}{\tan^2\beta(Z_{A_u})_{11}^2}
& \ \ {\rm up\ sector}.\end{array}\right.
    \end{split}
\eeq
The stronger bounds correspond to $x=1$ and phases $\lsim 0.3$. The
weaker bounds correspond to $x=4$ and phases $\lsim 0.1$. For
$\tan\beta\lsim 2(\gsim2)$ the down (up) sector represents the
stronger bound in Eq.~(\ref{eq:LRboundEDM}). The
$(\delta^{u}_{11})_{LR}$-related bound is largely insensitive to the
value of $\tan \beta$.

We also give the maximal values of the chirality-changing
$\delta^q_{LR}$ parameters in full anarchy using
Eq.~\eqref{eq:LRboundEDM} and $m_{\tilde q}=1$ TeV (for $i \neq j$):
\bea
(\delta^u_{ij})_{LR} &\lsim &6 \cdot 10^{-5}  , \nonumber \\
(\delta^d_{ij})_{LR} &\lsim &6 \cdot 10^{-5}/\tan \beta .
\eea

The constraints in Eq.~\eqref{eq:LRboundEDM} imply that, for
reasonable values of the RG-factor $r_3={\cal O}(1-10)$, most of the
$\delta^q_{LL}$ parameters are dominated by MFV effects:
\beqa\label{eq:delqlMFV}
(\delta^{d}_{12})_{LL} & \sim &
\frac{ c_u y_t^2|V_{ts}V_{td}^*|}{r_3}  ,\no \\
(\delta^{u}_{i3})_{LL}& \sim &
\frac{c_d y_b^2 |V_{ib}V_{tb}^*|}{r_3} ,\no\\
(\delta^{d}_{i3})_{LL}& \sim &
\frac{c_u y_t^2|V_{tb}V_{ti}^*|}{r_3} ,
\eeqa
where $i\neq 3$, whereas
\beqa\label{eq:delqlmaybeMFV}
(\delta^{u}_{12})_{LL} & \sim & \frac{\tilde r}{r_3} ,
\eeqa
with
\beq
\tilde r\equiv \max\{r, c_d y_b^2|V_{ub} V_{cb}^*|\}\sim \max\{r,
2\cdot 10^{-4} y_b^2\},
\eeq
can be either MFV or gravity dominated.  The
$\delta_{RR}^q$ parameters in full anarchy are still non-MFV
dominated, with the RR-mixing being subject to some subtleties
discussed below.  Order of magnitude estimates for the various
$\delta^q_{ij}$ are given in Table \ref{tab:delnoMFV}.

\begin{table}[t]
\caption{The order of magnitude estimates for
  $(\delta_{ij}^{d,u})_{LL,RR}$ and $\langle\delta^{d,u}_{ij}\rangle$ in
  the models defined by Eq. (\ref{eq:xzanar}). The numerical estimates are
  obtained using quark masses at the scale $m_Z$ \cite{Xing:2007fb}
  and for $r_3=3$ and $\tan \beta=3$. All numerical estimates scale as $(3/r_3)$.}
\label{tab:delnoMFV}
\begin{center}
\begin{tabular}{ll|ccc} \hline\hline
\rule{0pt}{1.2em}%
$q$ & $ij$\ &  $(\delta^q_{ij})_{LL}$ &  $(\delta^q_{ij})_{RR}$ &
$\langle\delta^q_{ij}\rangle$ \cr \hline
$d$ & $12$\ & $y_t^2|V_{ts}V_{td}^*|/r_3\sim 10^{-4}$ & $r/r_3\sim0.3r$
& $\sqrt{r y_t^2|V_{ts}V_{td}^*|}/r_3\sim 0.006\sqrt r$ \cr
$d$ & $13$\ & $y_t^2|V_{tb}V_{td}^*|/r_3\sim0.003$ &
$r/r_3\sim0.3 r$
& $\sqrt{r y_t^2|V_{tb}V_{td}^*|}/r_3\sim0.03\sqrt{r}$ \cr
$d$ & $23$\ & $y_t^2|V_{tb}V_{ts}^*|/r_3\sim0.01$ &
$r/r_3\sim0.3 r$
& $\sqrt{r y_t^2|V_{tb}V_{ts}^*|}/r_3\sim0.07\sqrt{r}$ \cr
$u$ & $12$\ & $\tilde r/r_3\sim0.3\tilde r$ & $r/r_3\sim0.3r$
& $\sqrt {r \tilde r}/r_3\sim0.3\sqrt{r \tilde r}$  \cr
$u$ & $13$\ & $y_b^2 |V_{ub}V_{tb}^*|/r_3\sim 4 \cdot 10^{-6}$ & $r/r_3\sim0.3r$
& $\sqrt{ r y_b^2 |V_{ub}V_{tb}^*|}/r_3\sim 0.001\sqrt{ r}$ \cr
$u$ & $23$\ & $y_b^2 |V_{cb}V_{tb}^*|/r_3\sim4 \cdot 10^{-5} $ & $r/r_3\sim0.3 r$
& $\sqrt{ r y_b^2 |V_{cb}V_{tb}^*|}/r_3\sim 0.004\sqrt{ r}$ \cr
\hline\hline
\end{tabular}
\end{center}
\end{table}

The situation regarding the RR flavor mixing is driven by three
factors: the extremely strong constraint on $r/r_3$, the fact that the
latter stems from LR-mixing, which is only suppressed by
$\sqrt{r/r_3}$, and the absence of significant MFV terms in the
RR-mixing as opposed to the LL one.

Concerning the MFV contributions in the RR sector, as implemented in
Eq.~(\ref{eq:delgen}), flavor-changing MFV effects in $(\widetilde
M^2_{q_R})_{ij}$ are absent at one-loop. However, they are generated
at two loops through the RGE \cite{Martin:1993zk}.  The largest such
effect is the contribution to $(\widetilde M^2_{d_R})_{23}$ and is
proportional to $y_s y_b y_t^2 V_{ts}^* V_{tb} \sim 2 \cdot 10^{-7}
\tan^2 \beta$. Estimating very roughly the loop suppression as $1/(16
\pi^2)^2 \times {\rm logs} \lsim 10^{-3}$, the non-MFV gravity effects
dominate in $(\delta^d_{23})_{RR}$ provided Eq.~\eqref{eq:LRboundEDM}
is saturated. We stress that in either case, the resulting
$\delta^q_{RR}$ value is very small and irrelevant for flavor
phenomenology. In this sense it is not of interest whether such a
small value is non-MFV or MFV. The latter can happen if the neutron
EDM bound tightens, or for very large values of $\tan \beta$.

Additionally, due to the smallness of the single insertion
$\delta^q_{RR}$ parameters, of order $10^{-7}$, the question arises as
to whether products of multiple mass insertions, effectively yielding
similar RR squark mixing $(\delta^{u,d}_{ij})_{RR}^{\rm eff}$, can
lead to comparable or larger effects.  These receive contributions
from $F$-terms induced by the supersymmetric Higgs mass term $\mu$ in
the MSSM superpotential.  For mixing involving the third generation we
estimate ($i=1,2$):
\bea
(\delta^{u}_{i3})_{RR}^{\rm eff}&=&(\delta^u_{i3})_{RL}
(\delta^u_{33})_{LR} \sim
\sqrt{\frac{r}{r_3} } \frac{m_t v \mu}{m_{\tilde q}^3}
\frac{1}{\sqrt{1+\tan^2 \beta}} \lsim   \frac{1 \cdot
  10^{-5}}{\tan\beta}\left(\frac{\mu}{1{\rm \ TeV}}\right),\no\\
(\delta^{d}_{i3})_{RR}^{\rm eff}&=&(\delta^d_{i3})_{RL}
(\delta^d_{33})_{LR} \sim \sqrt{\frac{r}{r_3} } \frac{m_b v
  \mu}{m_{\tilde q}^3} \frac{\tan \beta}{\sqrt{1+\tan^2 \beta}} \lsim
2 \cdot 10^{-7}\left(\frac{\mu}{1{\rm \ TeV}}\right) ,
\eea
where in the inequality we apply the bound Eq.~\eqref{eq:LRboundEDM}
for $\tan\beta\gsim2$ and use $m_{\tilde q}=1$~TeV.  For $\mu \sim
m_{\tilde q}$ we then find that the maximal reach of
$(\delta^{u,d}_{i3})_{RR}^{\rm eff}$ is larger than the corresponding
maximal reach of $(\delta^{u,d}_{i3})_{RR} \sim r/r_3$.  For all other $\delta^q$ parameters, the single mass insertion is the dominant one.

Combining the constraint Eq.~\eqref{eq:LRboundEDM} to the order of
magnitude estimates of Table \ref{tab:delnoMFV}, we obtain the upper
bounds on the $\delta^q_{ij}$ presented in Table \ref{tab:upbnoMFV}.
Note that in the up sector, entries with an $r_3$ dependence are MFV,
while non-$r_3$ entries are non-MFV.  As the
$(\delta^{u,d}_{i3})_{RR}$ and the $\langle \delta^{u,d}_{i3} \rangle$
parameters will not affect the maximal possible effects in the neutral
meson systems, we use the single mass insertions in both Tables
\ref{tab:delnoMFV} and \ref{tab:upbnoMFV}.

\begin{table}[t]
\caption{The order of magnitude upper bounds on
  $(\delta_{ij}^{d,u})_{LL,RR}$ and $\langle\delta^{d,u}_{ij}\rangle$
  for $r/r_3\lsim 1.2\cdot10^{-7}$, obeying the bound of
  Eq.~(\ref{eq:LRboundEDM}) for $\tan\beta=3$.
   Entries in parentheses are independent of
   $r$, therefore representing estimates rather than upper bounds.
}
\label{tab:upbnoMFV}
\begin{center}
\begin{tabular}{ll|ccc} \hline\hline
\rule{0pt}{1.2em}%
$q$ & $ij$\ &  $(\delta^q_{ij})_{LL}$ &  $(\delta^q_{ij})_{RR}$ &
$\langle\delta^q_{ij}\rangle$ \cr \hline
$d$ & $12$\ & $[10^{-4}(3/r_3)]$ & $1.2\cdot10^{-7}$ & $4\cdot 10^{-6}\sqrt{3/r_3}$ \cr
$d$ & $13$\ & $[0.003(3/r_3)]$ & $1.2\cdot10^{-7}$ & $2\cdot 10^{-5} \sqrt{3/r_3}$ \cr
$d$ & $23$\ & $[0.01(3/r_3)]$ & $1.2\cdot10^{-7}$ & $4\cdot 10^{-5}\sqrt{3/r_3}$ \cr
$u$ & $12$\ & $10^{-7}{\rm max}\{4.8/r_3,1.2\}$ & $1.2\cdot10^{-7}$ &
$10^{-7}{\rm \max}\{1.4\sqrt{3/r_3}, 1.2\}$  \cr
$u$ & $13$\ & $[4\cdot10^{-6}(3/r_3)]$ & $1.2\cdot10^{-7}$ &
$7\cdot10^{-7}\sqrt{3/r_3}$ \cr
$u$ & $23$\ & $[4\cdot10^{-5}(3/r_3)]$ & $1.2\cdot10^{-7}$ &
$2\cdot10^{-6}\sqrt{3/r_3}$ \cr
\hline\hline
\end{tabular}
\end{center}
\end{table}

We learn that the maximal possible effects in the neutral $B_d,\ B_s$
and $D$ systems are (for $r_3=3$ and, for the $D$ system, $\tan\beta =
3$):
\beq\label{eq:anarchy-A-reach}
\begin{array}{lc}
B_d: & |M_{12}^{\rm susy}/M_{12}^{\rm exp}|\lsim 2\cdot 10^{-4},\\
B_s: & |M_{12}^{\rm susy}/M_{12}^{\rm exp}|\lsim 5\cdot 10^{-4},\\
D: & |M_{12}^{\rm susy}/M_{12}^{\rm exp}|\lsim 3\cdot 10^{-10}. \end{array}
\eeq
The largest possible contributions to the $B_{d,s}$ mixing amplitudes
come from the MFV contributions to the $(\delta^d_{i3})_{LL}$'s.
Therefore, the effect is not enhanced for large $\tan\beta$.

We conclude that, for anarchical gravity-mediated contributions to the
$A$-terms, the effects on FCNC processes are negligibly small. In
contrast, any improvements in the neutron EDM measurements may either
further strengthen the constraints on this framework or discover its
effects.

\subsection{Yukawa-like $A$-terms}
\label{ssec:yukA}
Here we explore the implications of $A$-terms of Yukawa-like texture
as in the models with a FN symmetry discussed in Section
\ref{ssec:FCNCsumFN}, specifically, Eq.~(\ref{eq:yuklik}), and
\beqa\label{eq:Apreyuk}
(X_{q_A})_{ij}={\cal O}(1),\ \ \ (Z_{A_q})_{ij}\sim Y_{ij}^{q}\sim
V_{ij}{m_{q_j}}/{v_q}.
\eeqa

As concerns the $\delta^q_{LL,RR}$ parameters, the effect of such
$A$-terms can be described as ${\cal O}(r)$ changes in the RGE
coefficients $c_u, c_d, c_{uR}, c_{dR}$ defined in Eq.
(\ref{eq:msoftmz}); see Eqs.~(\ref{eq:mql2-soft}) and
\eqref{eq:mqr2-soft}. This leads, in turn, to at most ${\cal O}(1)$
changes in the $X_{q_A}$ matrices. Therefore, the estimates for the
$(\delta^q_{ij})_{LL,RR}$ parameters in this scenario vary by at most
${\cal O}(1)$ from the estimates obtained in Section \ref{ssec:noA}
for $Z_{A_q}=0$ at the high scale.

As concerns the $\delta^q_{LR}$ parameters, the parametric suppression
of $\widetilde A^{q}_{ij}$ can be extracted from $A^{q}_{ij}$ at the
high scale $m_M$.  This statement can be straightforwardly understood
in the up sector, regardless of the structure of the $A$-terms, and in
the down sector, for vanishing or anarchical initial $A$-terms. It is
a little more subtle for the down sector in the case of Yukawa-like
textured initial $A$-terms, but still holds true due the fact that the
various RGE contributions in the last two lines of Eq.
\eqref{eq:Aquarkmass-soft} are at most comparable to the direct term
proportional to $A^{d}_{ij}$.  For instance, in the
$(\delta^d_{LR})_{11}$ term, which is relevant for the EDM constraints,
$Y^d_{11}$ and $V_{td}^* Y^d_{31}$ are comparable.  Thus, the
expressions of \eqref{eq:delgen} for the $\delta^q$ parameters hold
with $Z_{A_q}$ of the structure \eqref{eq:Apreyuk}.

We conclude that the strongest constraint on $r/r_3$ in this scenario comes
from $\langle\delta^d_{12}\rangle$, as was the case for vanishing
$A$-terms.  Thus the bound of Eq.~\eqref{eq:rbound} holds, and
the estimates and
constraints that apply in the case of Yukawa-like $A$-terms are the
same as those that apply in the case of vanishing $A$-terms.

An exception to this is the prediction for the neutron EDM, which is induced by
the chirality-mixing $\delta^q_{LR}$ parameters in models with  texture
\eqref{eq:Apreyuk}:
\beq \label{eq:AYEDM}
|d_{n}^{\rm susy}/d_{n}^{\rm exp}|\lsim 0.01 \, (0.001) .
\eeq
Here, the first value corresponds to $x=1$ and a phase suppression in
$(\delta^q_{11})_{LR}$ of  $\sim 0.3$, and the value in parentheses
is obtained for  $x=4$ and a phase suppression of  $\sim 0.1$.

\subsection{Flavor constraints on the messenger scale}
\begin{figure}
\includegraphics[scale=1.0]{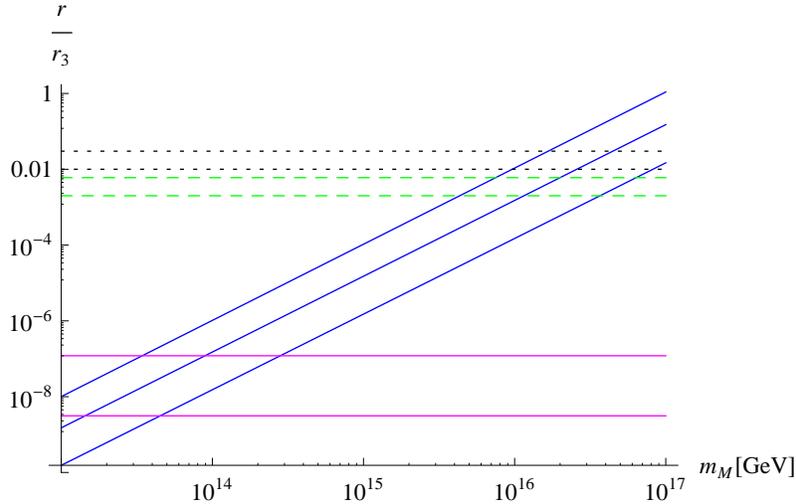}
\caption{\label{fig:MFVroverr3}
The three solid blue curves give $r/r_3$ as a function of the messenger
scale for $N_M=1$ (upper), $N_M=3$ (middle) and $N_M=10$
(lower). The three pairs of horizontal lines give upper bounds
on $r/r_3$ and correspond to the following flavor scenarios for the
gravity-mediated contributions: (i) Full anarchy (solid, pink) --
Eq.~(\ref{eq:LRboundEDM}) with $\tan\beta=3$; (ii) Anarchy with
vanishing or Yukawa-like $A$-terms  (dashed, green) -- Eq.~(\ref{eq:rbound}); (iii)
FN structure (dotted, black) -- Eq.~(\ref{eq:FNroverr3}). For a given
number of messengers and a given flavor scenario, FCNC and  CP constraints
give an upper bound on the messenger scale, which can be read from the
crossing point of the corresponding blue and horizontal curves.}
\end{figure}

The flavor and CP bounds on $r/r_3$, Eq.~(\ref{eq:FNroverr3}) or
(\ref{eq:rbound}) or (\ref{eq:LRboundEDM}), imply upper bounds on the
scale of gauge-mediation $m_M$ or, put differently, a minimal
separation between the scales of gauge- and gravity-mediation. In
minimal models \cite{Hiller:2008sv}, when the highest $F$-term
contributes to both gauge and gravity mediation,
\beq\label{eq:r-def}
r  \sim \left( \frac{m_M}{m_{\rm Pl}} \right)^2 \left( \frac{4
    \pi}{\alpha_3(m_M)} \right )^2 \frac{3}{8} \frac{1}{N_M} ,
\eeq
where $m_{\rm Pl}$ is the Planck scale and $N_M$ denotes the number of
messengers.  The dependence of $r_3$ on $m_M$ is only logarithmic, see
\cite{Hiller:2008sv} for details.  Fig.~\ref{fig:MFVroverr3} presents
the FCNC and CP constraints within the different flavor scenarios.

The difference between a FN-model (dotted lines) and models with
anarchy but vanishing or Yukawa-like $A$-terms (dashed lines) is
small, with the maximal messenger scales being related by $(m_M^{\rm
  max})_{A \sim0,Y}\sim\left(m_d/m_s\right)^{1/4} \, (m_M^{\rm
  max})_{FN}$.  In both cases,
\beq
m_M/m_{\rm Pl}\lsim10^{-3}\ \ \ ({\rm FN},\ {\rm or}\ {\rm anarchy}\
{\rm with} \ A\sim0, Y) .
\eeq
In the fully anarchical case, the EDM constraints yield a stronger
bound:
\beq
m_M/m_{\rm Pl}\lsim10^{-5}\ \ \ ({\rm full\ anarchy}) .
\eeq

\section{Conclusions}
\label{sec:con}
Measurements in low energy experiments of flavor-changing and of
CP-violating processes will complement direct searches for new
physics.  Our goal in this work has been to understand concrete ways
in which such complementarity will take effect. In order to do that,
we chose to work in a specific framework, which is supersymmetry where
the soft breaking terms receive dominant contributions from
gauge-mediation and sub-dominant contributions from gravity-mediation.
Within this framework, we investigated three main possible scenarios,
concerning the flavor structure of the gravity-mediated soft breaking
terms:
\begin{enumerate}
\item The flavor structure of all such terms is dictated by a
  Froggatt-Nielsen  symmetry, which is also responsible for the
  smallness and hierarchy in the Yukawa terms.
\item The flavor structure of the squark mass-squared matrix is
  anarchical, while the $A$-terms are subject to the FN selection rules.
\item The flavor structure of all soft supersymmetry breaking terms is
  anarchical.
\end{enumerate}
The first scenario has been studied in a previous paper
\cite{Hiller:2008sv}, while the latter two are explored in this work.
Additional scenarios that we investigated (one where holomorphic zeros
play a role and another where the $A$-terms vanish at the Planck scale)
provide further variations, but we will explain our main conclusions
on the basis of the three main ones.

Within this framework, we are able to answer the following questions,
which will be relevant whatever type of new physics will be
discovered:
\begin{itemize}
\item Which processes are most likely to show deviations from the
  standard model?
\item At what level can these effects appear?
\item Can we tell whether the flavor pattern of the gravity-mediated
  contributions is related to the standard model flavor pattern or not?
\end{itemize}

Table \ref{tab:con} should be helpful in clarifying our main
conclusions.
\begin{table}[t]
\caption{The maximal size of possible effects in the mixing of
  $B_d-\overline B_d$, $B_s-\overline B_s$ and $D^0-\overline D^0$ for
  low $\tan \beta$ [Eqs.~(\ref{eq:FN-reach}),
  (\ref{eq:anarchy-reach}), (\ref{eq:anarchy-A-reach})], normalized to
  the experimental value, and in the neutron EDM $d_n$
  [Eqs.~(\ref{eq:FNEDM}), (\ref{eq:AYEDM})], normalized to the
  experimental upper bound. FN (An) means that the structure of
  the corresponding soft terms defined in Eqs.~(\ref{eq:inicon}) and
  (\ref{eq:iniconA}) is
  dictated by a Froggatt-Nielsen symmetry (is anarchical).
}
\label{tab:con}
\begin{center}
\begin{tabular}{c c|c c c |c} \hline\hline
\rule{0pt}{1.2em}%
$X$ & $Z$ & $B_d$ & $B_s$ & $D^0$ & $d_n$ \cr \hline
 FN & FN  & 0.002 & 0.005 & 0.03  & 0.02 \cr
 An & FN  & 0.007 & 0.002 & ${\cal{O}}( 1)$   & 0.01 \cr
 An & An  & $2 \cdot  10^{-4}$ & $5 \cdot 10^{-4}$ & ${\cal{O}}( 10^{-10})$ & 1 \cr
 \hline \hline
\end{tabular}
\end{center}
\end{table}

We learn that any improvement in the upper bound on CP violation in
neutral $D$ mixing or on the neutron EDM will have an impact on our
framework. It will make the upper bound on the ratio of
gravity-to-gauge mediation stronger (in, respectively, the An-FN and
An-An scenarios) or, equivalently, it will strengthen the upper bound
on the messenger scale of gauge-mediation. Conversely, if non-MFV
effects are observed, their size will provide a lower bound on the
size of gravity-mediated contributions. In models of a single scale
supersymmetry breaking, such a bound can be translated into a lower
bound on this scale.

We can further make the following general statements, some of which
are valid well beyond the specific framework of new physics that we
have studied:
\begin{itemize}
\item If the flavor and CP structure of the new physics is anarchical,
  it is quite possible that its effects will be discovered in
  CP-violating flavor-diagonal observables rather than in
  flavor-changing measurements.
\item If new flavor effects of similar relative sizes are discovered
  in both third generation ($b$ or $t$) decays and second generation
  ($c$ decays), then the new physics is likely to have a flavor
  structure. Furthermore, this structure may well be related to the
  standard model one.
\item A situation where non-MFV effects related to third generation
  physics are larger (in their relative size) than those related to
  charm physics requires that the flavor structure of the new physics
  is not related to the standard model one.
\end{itemize}

It is amusing to note that since the strong suppression of gravity-mediated flavor
within full anarchy  makes the latter almost MFV-like, it allows
for the possibility of a long-lived light stop with macroscopic decay
length given a suitable particle spectrum \cite{Hiller:2008wp}.

We conclude that the search for new physics in high-precision flavor
experiments and in EDM measurements will be very informative about the
underlying, very high scale new physics. It will further allow us to
test whether mechanisms such as a Froggatt-Nielsen symmetry dictate
all flavor structures or not.

\acknowledgments
This work was supported by a grant from the G.I.F., the
German--Israeli Foundation for Scientific Research and Development,
and by the Minerva Foundation. The work of YN is supported in part
by the United States-Israel Binational Science Foundation (BSF),
and by the Israel Science Foundation (ISF).

\appendix
\section{The high scale-low scale connection
\label{sec:rge}}
In supersymmetric models with hybrid gauge- and gravity-mediation of
supersymmetry breaking, the form of the soft terms at the scale of
gauge-mediation $m_M$ is given in Eqs.~(\ref{eq:inicon}) and
(\ref{eq:iniconA}). In this Appendix, we consider the RGE effects and
provide approximate analytical expressions for the soft terms at the
weak scale, $m_Z$.

\subsection{Weak scale}
The weak-scale squark mass-squared matrices $\overline M^2_{\tilde
  q_A}$ have the following form:
\beqa\label{eq:squark-masses}
\overline M^2_{\tilde D_L}&=&M ^2_{\tilde Q_L}+D_{D_L}{\bf 1}
+m_Dm_D^\dagger ,\no\\
\overline M^2_{\tilde U_L}&=&M^2_{\tilde Q_L}+D_{U_L}{\bf 1}
+m_Um_U^\dagger ,\no\\
\overline M^2_{\tilde D_R}&=&M^2_{\tilde D_R}+D_{D_R}{\bf 1}
+m_D^\dagger m_D ,\no\\
\overline M^2_{\tilde U_R}&=&M^2_{\tilde U_R}+D_{U_R}{\bf 1}
+m_U^\dagger m_U,
\eeqa
where $m_{U,D}$ are the up and down quark mass matrices in the flavor
basis, $D_{q_A}$ are the $D$-term contributions, and all quantities
should be evaluated at the electroweak scale $\mu \sim m_Z$.

The initial conditions Eqs.~(\ref{eq:inicon}) and (\ref{eq:iniconA})
hold at the scale of gauge-mediation $m_M$, while flavor-changing
processes restrict the weak scale parameters $(\delta^q_{ij})_{NM}$
($N,M=L,R$), and so the effect of RG evolution on the soft terms must
be taken into account.

As concerns the soft squark masses, we obtain
the following approximate form:
\beqa\label{eq:msoftmz}
M^2_{\tilde Q_L}(m_Z)& \sim&\tilde m^2_{Q_L} (r_3 {\bf 1}+
c_u Y_u Y_u^\dagger + c_d Y_d Y_d^\dagger+ r X_{q_L} + r Z_{A_u} Z_{A_u}^\dagger
+ r Z_{A_d} Z_{A_d}^\dagger), \nonumber\\
M^2_{\tilde U_R}(m_Z)&\sim&\tilde m^2_{U_R} (r_3 {\bf 1}+
c_{uR} Y_u^\dagger Y_u+ r X_{u_R}+ r Z_{A_u}^\dagger  Z_{A_u}), \nonumber \\
M^2_{\tilde D_R}(m_Z)& \sim&\tilde m^2_{D_R} (r_3 {\bf 1}+
c_{dR} Y_d^\dagger Y_d + r X_{d_R}+r Z_{A_d}^\dagger Z_{A_d}),
\eeqa
where $Y_u$ and $Y_d$ denote the up and down quark Yukawa matrices in
the flavor basis, and the effect of trilinear soft couplings is
included. (The RG-coefficient in front of the $A$-terms is of order
one.) We note the following points:
\begin{enumerate}
\item The factor $r_3$ captures the effect of RGE corrections to the
  diagonal elements of the soft squark mass matrices
  $(M^2_{\tilde q_A})_{ii}$ via
\beq\label{defrthree}
\tilde m^2_{12}(\mu=m_Z)=r_3\tilde m^2_{12}(\mu=m_M) ,
\eeq
with the average diagonal mass-squared defined as
\beq \label{eq:m12tilde}
\tilde m^2_{ij} \equiv \frac{1}{2}
\left( (M^2_{\tilde q_A})_{ii}+(M^2_{\tilde q_A})_{jj} \right) .
\eeq
For simplicity this factor is taken to be universal among all squarks,
as the dominant contribution to the initial squark soft masses and to
their RGE is QCD-induced and, in the limit that we neglect the
electroweak gauge couplings, is indeed universal.
\item
The coefficients $c_u,c_d,c_{uR},c_{dR}$ can be of ${\cal{O}}(1)$ for
$m_M \sim m_{\rm GUT}$ and are all negative. Yukawa-related
contributions to the RGE thus lower the weak scale values of the
diagonal $(M^2_{\tilde q_A})_{33}$ entries with respect to the high
scale ones. Sub-dominant MFV terms with higher powers of the Yukawa
couplings are henceforth neglected.
\item We use the various $\tilde m_{ij}^2(m_Z)$ to
evaluate the denominator of the $(\delta^q_{ij})_A$, neglecting
$D$-terms of ${\cal{O}}(m_Z^2/\tilde m_{ij}^2)$ and $F$-terms of at
most ${\cal{O}}(m_t^2/\tilde m_{i3}^2)$.
\item
Eq.~(\ref{eq:msoftmz}) is given in the flavor basis, while the
$\delta^q$ parameters are relevant in the mass basis of the quarks.
This rotation of the squarks leaves the parametric pattern of the
$X_{ij}$ and $Z_{ij}$ unchanged.
\end{enumerate}

As concerns the $A$-terms, we obtain for small to moderate $\tan\beta$
(so that for the purpose of RGE we employ $Y_d \ll 1$) the following
approximate form:
\beqa\label{eq:Asoftmz}
A^u(m_Z)& \sim & M_3 (a_u + b_u Y_u Y_u^\dagger) Y_u +
\tilde m \sqrt{r} (d_u+ e_u Y_u Y_u^\dagger) Z_{A_u}  ,\no\\
A^d(m_Z)& \sim & \tilde m  \sqrt{r}  (d_d+ e_d Y_u Y_u^\dagger) Z_{A_d} ,
\eeqa
where $\tilde m$ represents the typical scale of the $SU(3)$-related
contributions to the gauge-mediated soft masses. The $a,b$
coefficients relate to the MFV part of the trilinear scalar couplings
and are irrelevant to our discussion.  The dimensionless coefficients
$d,e$ can be ${\cal O}(1)$ and, for our purposes, are taken as such.

\subsection{The $\delta^q$ parameters}
We now work in the basis in which the quarks mass matrices and gluino
couplings are diagonal. We use $q=U,D$, $i =1,2$, $j=1,2,3$, $r \ll
r_3$ and neglect the masses of the first and second generation quarks.
For the LL block, we find
\beqa\label{eq:mql2-soft}
(\widetilde M^2_{\tilde q_L}(m_Z))_{33}& \sim &
\tilde m^2_{Q_L} (r_3 + c_u y_t^2 + c_d y_b^2
+r (X_{q_L})_{33}+r (Z_{A_u} Z_{A_u}^\dagger)_{33}
+ r (Z_{A_d} Z_{A_d}^\dagger)_{33}), \nonumber \\
(\widetilde M^2_{\tilde q_L}(m_Z))_{ii}& \sim &
\tilde m^2_{Q_L} (r_3 +r (X_{q_L})_{ii}+c_u y_t^2|V_{ti}|^2
+r (Z_{A_u} Z_{A_u}^\dagger)_{ii}
+ r (Z_{A_d} Z_{A_d}^\dagger)_{ii}) , \nonumber \\
(\widetilde M^2_{\tilde U_L}(m_Z))_{12}& \sim &
\tilde m^2_{Q_L} (c_d y_b^2 V_{ub} V_{cb}^*+r (X_{u_L})_{12}
+r (Z_{A_u} Z_{A_u}^\dagger)_{12}
+ r (Z_{A_d} Z_{A_d}^\dagger)_{12}) ,\nonumber \\
(\widetilde M^2_{\tilde U_L}(m_Z))_{i3}&\sim &
\tilde m^2_{Q_L} (c_d y_b^2 V_{ib} V_{tb}^*+r (X_{u_L})_{i3}
+r (Z_{A_u} Z_{A_u}^\dagger)_{i3}
+ r (Z_{A_d} Z_{A_d}^\dagger)_{i3}) ,\nonumber \\
(\widetilde M^2_{\tilde D_L}(m_Z))_{12}& \sim &
\tilde m^2_{Q_L} (c_u y_t^2 V_{ts} V_{td}^*+r (X_{d_L})_{12}
+r (Z_{A_u} Z_{A_u}^\dagger)_{12}
+ r (Z_{A_d} Z_{A_d}^\dagger)_{12}) ,\nonumber \\
(\widetilde M^2_{\tilde D_L}(m_Z))_{i3}& \sim &
\tilde m^2_{Q_L} (c_u y_t^2 V_{tb} V_{ti}^*+r (X_{d_L})_{i3}
+r (Z_{A_u} Z_{A_u}^\dagger)_{i3}
+ r (Z_{A_d} Z_{A_d}^\dagger)_{i3}).
\eeqa
For the RR block, we find
\beqa\label{eq:mqr2-soft}
(\widetilde M^2_{\tilde U_R}(m_Z))_{33}& \sim &
\tilde m^2_{U_R} (r_3 + c_{uR} y_t^2 +r (X_{u_R})_{33}
+r (Z_{A_u}^\dagger  Z_{A_u})_{33}), \nonumber \\
(\widetilde M^2_{\tilde D_R}(m_Z))_{33}& \sim &
\tilde m^2_{D_R} (r_3 + c_{dR} y_b^2 +r (X_{d_R})_{33}
+r (Z_{A_d}^\dagger  Z_{A_d})_{33}), \nonumber \\
(\widetilde M^2_{\tilde q_R}(m_Z))_{ii}& \sim &
\tilde m^2_{q_R} (r_3 +r (X_{q_R})_{ii}
+r (Z_{A_q}^\dagger  Z_{A_q})_{ii}) , \nonumber \\
(\widetilde M^2_{\tilde q_R}(m_Z))_{ij}& \sim&
\tilde m^2_{q_R}( r (X_{q_R})_{ij} +
r (Z_{A_q}^\dagger  Z_{A_q})_{ij}), ~~~(i \neq j),
\eeqa
For the LR block, we find ($i,j\neq 3$):
\beqa\label{eq:Aquarkmass-soft}
(\widetilde A^{u}(m_Z))_{33}& \sim & M_3 (a_u+b_u y_t^2) y_t +\tilde m
\sqrt{r} (d_u+e_u y_t^2) (Z_{A_u})_{33} , \nonumber \\
(\widetilde A^{u}(m_Z))_{ij}& \sim & \tilde m
\sqrt{r}  d_u (Z_{A_u})_{ij} , \nonumber \\
(\widetilde A^{u}(m_Z))_{3i}& \sim & \tilde m
\sqrt{r}  (d_u +e_u y_t^2)(Z_{A_u})_{3i} , \nonumber \\
(\widetilde A^{d}(m_Z))_{33}& \sim & \tilde m
\sqrt{r} (d_d+e_d y_t^2) (Z_{A_d})_{33} , \nonumber \\
(\widetilde A^{d}(m_Z))_{3j}& \sim & \tilde m\sqrt{r}
\left(d_d (Z_{A_d})_{3j} + e_d y_t^2 V_{tb}^*  V_{tk} (Z_{A_d})_{kj}  \right),  \nonumber \\
(\widetilde A^{d}(m_Z))_{ij}& \sim & \tilde m\sqrt{r}
\left(d_d (Z_{A_d})_{ij}+e_d y_t^2 V_{ti}^*V_{tk} (Z_{A_d})_{kj}  \right) .
\eeqa
(In the last two lines summation over $k=1,2,3$ is implied.)  In some
specific cases, for example, vanishing initial $A$-terms or anarchical ones,
the above expressions simplify.

Eqs. (\ref{eq:mql2-soft}), (\ref{eq:mqr2-soft}) and
(\ref{eq:Aquarkmass-soft}) lead to the expressions for the $m_Z$-scale
flavor-changing $\delta^q$ parameters given in Eq.~(\ref{eq:delgen}).

A final comment here concerns $\delta^u_{LR}$. Here, there are
contributions from both gravity-mediation (via the initial $A$-terms)
and gauge-mediation (via the gaugino masses).  We neglect the gaugino
contributions when constraining the low energy $\delta^u_{LR}$, since we
are interested in the constraints on the gravity parameters.
Constraints from the gaugino contributions are similar to those in
gauge-mediation and do not concern us here.


\end{document}